\documentclass[12pt]{elsarticle}

\usepackage[utf8]{inputenc}
\usepackage[english]{babel}

\usepackage{url}

\usepackage{enumitem}
\usepackage{amsfonts}
\usepackage{amssymb}
\usepackage{graphicx}
\usepackage{color} 
\usepackage{lscape}
\usepackage{caption}
\usepackage{etoolbox} 
\usepackage{booktabs}
\robustify\bfseries

\usepackage{tabularx}
\newcolumntype{L}[1]{>{\raggedright\let\newline\\\arraybackslash\hspace{0pt}}m{#1}}
\newcolumntype{C}[1]{>{\centering\let\newline\\\arraybackslash\hspace{0pt}}m{#1}}
\newcolumntype{R}[1]{>{\raggedleft\let\newline\\\arraybackslash\hspace{0pt}}m{#1}}

\begin{document}
\title{Measuring breathing induced oesophageal motion and its dosimetric impact}

\author[1,2]{Tobias Fechter\corref{cor1}}
\ead{tobias.fechter@uniklinik-freiburg.de}

\author[2,3]{Sonja Adebahr}

\author[2,3]{Anca-Ligia Grosu}

\author[1,2]{Dimos Baltas}

\cortext[cor1]{Corresponding author}


\affiliation[1]{Division of Medical Physics, Department of Radiation Oncology, Medical Center – University of Freiburg, Faculty of Medicine. University of Freiburg, Germany}
\affiliation[2]{German Cancer Consortium (DKTK). Partner Site Freiburg, Germany}
\affiliation[3]{Department of Radiation Oncology, Medical Center – University of Freiburg, Faculty of Medicine. University of Freiburg, Germany}


\begin{abstract}
\textbf{Purpose:} Stereotactic body radiation therapy allows for a precise dose delivery. Organ motion bears the risk of undetected high dose healthy tissue exposure. An organ very susceptible to high dose is the oesophagus. Its low contrast on CT and the oblong shape render motion estimation difficult. We tackle this issue by modern algorithms to measure oesophageal motion voxel-wise and estimate motion related dosimetric impacts.
 
\textbf{Methods:} Oesophageal motion was measured using deformable image registration and 4DCT of 11 internal and 5 public datasets. Current clinical practice of contouring the organ on 3DCT was compared to timely resolved 4DCT contours. Dosimetric impacts of the motion were estimated by analysing the trajectory of each voxel in the 4D dose distribution. Finally an organ motion model for patient-wise comparisons was built.
  
\textbf{Results:} Motion analysis showed mean absolute maximal motion amplitudes of 4.55 $\pm$ 1.81 mm left-right, 5.29 $\pm$ 2.67 mm anterior-posterior and 10.78 $\pm$ 5.30 mm superior-inferior. Motion between cohorts differed significantly. In around 50 $\%$ of the cases the dosimetric passing criteria was violated. Contours created on 3DCT did not cover 14 $\%$ of the organ for 50 $\%$ of the respiratory cycle and were around 38 $\%$ smaller than the union of all 4D contours. The motion model revealed that the maximal motion is not limited to the lower part of the organ. Our results showed motion amplitudes higher than most reported values in the literature and that motion is very heterogeneous across patients. 

\textbf{Conclusions:} Individual motion information should be considered in contouring and planning.
\end{abstract}


\begin{keyword}
Intra-fraction motion \sep Oesophagus \sep Lung cancer \sep SBRT \sep 4DCT \sep 4D Dose \sep Deformable image registration
\end{keyword}

\maketitle

\section{Introduction}

Modern stereotactic body radiation therapy (\textit{SBRT}) in the thoracic region allows for a precise and accurate dose delivery. The steep dose gradients that are possible with SBRT facilitate a dose boost in the target volume while sparing the organs at risk (\textit{OARs}). However, during treatment it is possible that certain OARs move e.g. due to respiration, heart beat, swallowing or intrinsic movements which can cause OAR-exposure to the high dose field. An OAR that requires special care when treating tumours in or close to the mediastinal region is the oesophagus. Its radiosensitive mucosa making it susceptible to injuries due to higher dose exposure. Thus, sometimes severe sequeales as oesophagitis, hemorrhagia, fistula with mediastinitis or strictures can occur. Consequently great caution has been exercised implementing SBRT to central regions, especially to lesions in proximity to the oesophagus \cite{Timmerman2006,adebahr2015}, as  for SBRT  - applying huge biological effective dose to the tissue -  mediastinal tolerance doses are not known. Several retrospective data suggest dosimetric constraints (e.g. \citet{adebahr2015}). However, it is not clear what dose really leads to severe harm to the oesophagus. Thus, it is necessary to keep the oesophagus out of the high dose irradiation field. Therefore a precise definition of the organ's boundaries and an estimation of its motion is required.


The current approach to demarcate the oesophagus for treatment planning is to delineate its outline on ungated computed tomography (\textit{CT}) or on average CT images (computed on basis of 4DCT) \cite{kong2011}. This procedure has two drawbacks. First, a precise organ outline is not possible due to the blurriness of average CT or the motion artifacts in ungated CT. Second, ungated CT does not allow for an estimation of organ motion at all, on average CT the motion can only be guessed from the motion blur. The motion information from a time gated CT acquisition (\textit{4DCT}) is usually considered only for the target volume itself but neglected for OARs as it would be an elaborate and time consuming procedure. 

Motion estimation of the oesophagus has been the scope of several publications with a multitude of applied techniques and heterogeneous results. For the alignment of 4DCT phase images rigid \cite{Gao2019,Kobayashi2016,cohen2010,Abhijit2009} as well as deformable image registration (\textit{DIR}) \cite{Palmer2014,Yaremko2008} was utilised. \citet{Sekii2018} and \citet{doi2018} used fiducial markers for improving accuracy. Another point that differs between the works published so far, is the analysed part of the organ. \citet{Gao2019} and \citet{Kobayashi2016} investigated the shift of the oesophageal centroid. \citet{cohen2010} and \citet{doi2018} divided the organ in a upper and lower region whereas \citet{Sekii2018} analysed the motion separately for upper, middle and lower part. \citet{Palmer2014} focused on heart beat induced motion and \citet{Yaremko2008} and \citet{WEISS200844} investigated the motion around the tumour. The presented motion amplitudes range from below 1 mm in all 3 dimension \cite{Gao2019} to 4.1 mm anterior-posterior (\textit{AP}) \cite{Palmer2014}, 4.1 mm left-right (\textit{LR}) \cite{cohen2010} and 8.0 mm superior-inferior (\textit{SI}) \cite{Abhijit2009}. 
We attribute the heterogeneous results to a number of points: Rigid registration can not depict the complex nature of the mediastinal soft tissue motion. Although the usage of fiducials can give very precise results for the location where the fiducial is placed, the deformation between fiducials needs to be estimated. Additionally, the different analysed regions of the organ hamper a direct comparison of the published values.

That oesophageal motion causes a deviation from the planned dose, but only within the constraints was shown by \citet{Wang2019} and \citet{EHRBAR201645}. \citet{Chung2018} reconstructed the actual delivered dose for 30 gated SBRT patients. With a rigid motion model for the whole patient, they found no clinically significant dose deviations but that gamma passing rates decrease with an increase in the motion amplitude and can be violated for certain lung patients. \citet{pmid18722278} investigated the difference in dose parameters between 3DCT and 4DCT based treatment plans for oesophageal cancer patients. Their volume based analyses showed that plans based on a 3DCT scan could overestimate the target coverage.

In this work we present a method to estimate the respiration induced oesophageal motion voxel-wise for the whole organ. Then we investigate the current clinical practice to contour the oesophagus on 3DCT and if this technique is prone to missing parts of the organ or overestimating the volume. In a third step the motion related dosimetric impact is assessed by evaluating tissue motion in a 4D dose distribution. Lastly, we build a motion model of the oesophagus allowing for a patient wise motion comparison and an illustration of the motion not limited to predefined regions.

\section{Data}

Oesophageal motion was analysed for 11 internal (\textit{C1 - C11}) datasets from the department of radiation oncology, university medical centre, Freiburg and 5 external (\textit{T1 - T5}) datasets from The Cancer Imaging Archive. Each of them consisting of a 4D retrospectively respiratory gated CT acquisition consisting of 10 3D phase images. In addition the 11 internal datasets had an average CT and information about the delivered dose in terms of 3D and 4D dose volumes.
9 of the 16 datasets depicted the whole organ, 3 showed the oesophagus in the total lung region, on 2 datasets the central part was visible, 1 dataset showed the lower and 1 dataset the upper part of the oesophagus. The minimum depicted length was 180 mm.

\subsection{Internal Datasets}
The 11 clinical datasets were acquired on a Gemini TF Big Bore or Gemini Brilliance Big Bore (Philips Healthcare, Andover, MA, USA), with voxel resolution 0.97 $\times$ 0.97 mm - 1.17 $\times$ 1.17 mm in axial plane and 2.00 mm in z-direction. Volumes had 90 - 114 slices, each slice of size 512 $\times$ 512 voxels. All patient suffered from central non-small-cell lung cancer (\textit{NSCLC}), received 60 Gy in 8 fractions of 7.5 Gy. Dose prescription was chosen such that 95 \% of the PTV received at least the nominal fraction dose, and 99 \% of the PTV received a minimum of 90 \% of the nominal dose (according to ICRU Report 83 \cite{ICRUReport83}). For dosimetric analysis we used 11 dose volumes per patient. The baseline dose, which represents current clinical practice, was determined by calculating the delivered dose on the basis of average CT (\textit{3DDose}). In addition, to measure the impact of tissue motion and the associated change in physical properties, the applied dose was recalculated (with same beam configuration and without new optimisation) for each time phase CT separately creating a 4D dose volume (\textit{4DDose}). Each time-phase dose representing a tenth of the total dose \cite{Rouabhi2015}. The temporal schedule of dose delivery was neglected. Dose calculations were done with Eclipse treatment planning software v.15.6 (Varian Medical Systems, Palo Alto, CA, USA). The study was approved by the local ethics committee. 

\subsection{The Cancer Imaging Archive Datasets}
The 5 external datasets were retrieved from the 4D-Lung collection \cite{Hugo2016} available at The Cancer Imaging Archive (\textit{TCIA}) \cite{Clark2013}. The 4D-Lung collection provides locally advanced NSCLC patient datasets. All taken datasets consisted of a 4D fan beam CT scan acquired on a 16-slice helical CT scanner (Brilliance Big Bore, Philips Healthcare, Andover, MA, USA) divided into 10 breathing phases and oesophagus contours for each time phase created by a single Radiation Oncologist. For our analysis we used the patients 100, 101, 102, 103 and 107. Only for those patients 4D oesophagus contours were available. Voxel spacing was 0.97 $\times$ 0.97 $\times$ 3 mm in x-, y- and z-direction. Each CT slice consisted of 512 $\times$ 512 voxels. The number of slices ranged from 84 to 149. 

\section{Methods}

To accomplish the goals defined in the introduction we needed to preprocess the datasets, then identify the oesophagus on each time phase and average CT scan by segmenting the images and eventually calculating trajectories by DIR. 

\subsection{Preprocessing}

Image registration of 4D datasets is a very elaborate task. To reduce the amount of data to be processed every dataset was cropped to a region of interest (\textit{ROI}). After the oesophagus contours were completed the rectangular cuboid shaped ROI was defined as follows: combine all time phase contours and the average CT contour with a logical OR operator to a union contour, subsequently determine the minimum and maximum positions of the union contour in x-, y- and z-direction, finally add a margin of 20 voxels on each side. In the course of this work we tried different margins. A margin of 20 voxels revealed to be a good trade-off between including all relevant structures and faster image processing. For dosimetric analysis we re-sampled the dose to image resolution with Plastimatch v.1.7 \cite{plastimatchWeb} and tri-linear interpolation. 

For the creation of the inter patient motion model the maximum inhale datasets were re-sampled to the resolution of one reference dataset (Case 100 of TCIA) after affine registration with elastix's \cite{elastix2019} third order B-spline interpolation.

\subsection{Segmentation}

The oblong shape and the poor contrast to surrounding tissue render the delineation of the oesophagus a time-consuming task. Therefore, we employed our recently developed CNN-based algorithm for automatic oesophagus delineation \cite{Fechter2017} to create an initial contour on each 3D time phase image and the average CT of the internal datasets. The generated contours where then checked and manually corrected by one observer and double-checked by an experienced radiation oncologist. The final 121 reference contours follow the EORTC 22113-08113 Lungtech protocol and guidelines \cite{kong2011,adebahr2015}. For the external datasets the contours provided by TCIA were used. Manual contouring and visual inspection was done with 3D Slicer v.4.10.0 \cite{FEDOROV20121323}.

\subsection{Registration}

In this project inter-patient and intra-patient registration were conducted. Intra-patient registration was used to measure the oesophageal motion for each patient and inter-patient registration to create the motion model. In a 4DCT the time phase images are already rigidly aligned by the scanner therefore solely DIR was needed for the intra-patient registration, whereas for inter-patient alignment affine and deformable registration were necessary.

Elastix \cite{elastix2019} was used for affine image registration with the settings provided for CT lung registration with mutual information (Par0003 available online \cite{elastixParamsWeb} presented in the work by \citet{elastix2019}). DIR was done with the 4D algorithm by \citet{Fechter2020} which is able to consider the underlying cyclic respiratory motion pattern. In this work we further extended the algorithm by a diffeomorphic layer for the reduction of anatomically implausible deformations and the option to make use of segmentation information to steer the registration process. The code is available online \cite{oneShotWeb}.

For the intra-patient registration we used solely CT data and no contour information. Preliminary experiments showed that using the contour information reduces the accuracy due to slight contour inconsistencies. The inter-patient registration is more challenging due to anatomical variations between patients. Therefore we had to make use of CT as well as contour information (in terms of distance maps) for DIR. For the patients C4 and C10 acceptable results could only be generated with contour based information and neglecting CT volume information because of anatomical differences.

After a general evaluation (see \ref{par:regQA}) of the registration algorithm, the quality of inter- and intra-patient registration was evaluated manually by one experienced observer considering anatomical landmarks. Additionally, we calculated the Jacobian matrix for every point in
a deformation vector field (\textit{DVF}) to detect anatomically implausible deformations.

\subsection{Evaluation}

To evaluate the similarity between contours we considered 3 methods. Volume similarities were measured by the Sørensen-Dice index (\textit{DSC}) \cite{sorensen1948method}. Volume-based metrics alone might miss clinically relevant differences as  they  show  a  lower  sensitivity  to  errors where outlines deviate and the volume of the erroneous region is  small  compared  to  the  total  volume.  Thus,  we  considered also  distance-based  metrics  like  the  Hausdorff  distance  (HD) and the average symmetric surface distance (ASSD). DSC, HD and ASSD were calculated using MedPy v.0.4.0 \cite{medPyWeb}.

To compare dose volumes we used the gamma analysis \cite{low1998} provided by Plastimatch v.1.7. The gamma criterion at a given position is calculated by comparing the dose value in the reference dose volumes to the dose values within a given range in the other dose volumes. If the gamma value is below or equal to one the dose values at a given position are equal or differ by an acceptable amount in a given range. In our experiments we calculated the local gamma index in two different ways: First, we followed the recommendations by the AAPM \cite{miften2018} with a passing criteria of $3 \% \// 2 mm$, a low dose cut-off of 10 \% of the prescription dose (60 Gy) and a minimum passing rate of 90 \% for acceptable deviations. As, for some of our cases, the oesophagus resides in the low dose area and a big part is cut-off with the 6 Gy threshold, we performed a second gamma experiment with weaker criteria of $3 \% \// 3 mm$ as it is common in the literature \cite{HUSSEIN20171,Han2018}, a low dose cut-off of 10 \% of the maximum dose inside the oesophagus and a minimum passing rate of 80 \% \cite{Han2018}. In addition to the gamma analyses we provide the dosimetric indices $D_{max}$, $D_{mean}$, $D_{2\%}$, $D_{33\%}$ and $D_{50\%}$ for the oesophagus. $D_{max}$ is the maximal dose the organ receives, $D_{x\%}$ ($x \in \{2, 33,50\}$) gives the dose received by at least $x \%$ of the volume.

Statistical   analysis   was   performed   with   the   Wilcoxon signed-rank test. This test, which has for null hypothesis that the median group difference is zero, was chosen due to non-normal  distribution  and  heteroscedasticity  of  the  data.  In  our experiments, the confidence alpha was set to 5 \%.

\subsection{Mathematical Notation}
\label{subsec:ImgDoseProc}

Let $I^N$ be a 4D image dataset consisting of $N$ 3D images. In this work three types of image datasets were used: CT volumes, dose volumes and contours in binary image representation. The DIR algorithm yields a transformation $T^N$ that maps each 3D image $I_n$ in $I^N$ to its timely adjacent successor $I_{n+1}$ (as we deal with periodic data $I_{N-1}$ is aligned to $I_0$). $T^N$ consists of $N$ 3D dense vector fields and describes the trajectory for each voxel in $I^N$. $T_i^j(x)$ is the sequential application of the deformation fields $T_i$ to $T_j$ to the position of voxel $x$. 

\section{Results}

\subsection{Registration QA}
\label{par:regQA}

To ensure that the extended version of the used DIR algorithm is as accurate as the published version \cite{Fechter2020} we applied the algorithm to the maximum inhale and exhale CT scans of publicly available POPI and DirLab datasets \cite{vandemeulebroucke2011spatiotemporal,Castillo2009a,Castillo2009}. Both provide manually set landmarks for all CT scans. We conducted the evaluation by comparing the landmark distances after the application of the original and the extended DIR algorithm. In addition, to ensure that the registration results of aligning the maximum inhale and exhale phase (\textit{3D registration}), which was done for the POPI and DirLab experiment, are comparable to the alignment of all consecutive time phases in a 4D dataset (\textit{4D registration}), we used the calculated DVFs to deform the oesophagus contours of the maximum inhale to the maximum exhale phase and vice versa. First, with the 4D DVF and then with the 3D DVF. To deform a contour $I_i$ of time phase $i$ to phase $j$, the DVFs from $T_i$ to $T_j$ are applied sequentially $T_i^j(I_i)$. Registration quality can then be measured by calculating DSC, HD and ASSD between $T_i^j(I_i)$ and $I_j$.

The registration of the POPI and DirLab datasets reduced the average registration error $\pm$ standard deviation (\textit{SD}) from 1.70 $\pm$ 2.15 mm to 1.39 $\pm$ 1.18 mm compared to our previous work. The positive impact of the algorithm changes made the new version eligible for usage in this work. The experiment for comparing 3D and 4D registration showed that the 4D registration (median DSC, HD and ASSD of 0.84, 7.78 mm and 0.94 mm) yielded slightly worse but still reasonable results compared to the 3D registration (median DSC, HD and ASSD of 0.85, 6.05 mm and 0.86 mm).

\subsection{Comparison of Contouring Methods}
\label{subsec:ComparisonofContouringMethods}
\begin{figure}[htbp] 
  \centering
     \includegraphics[width=\textwidth]{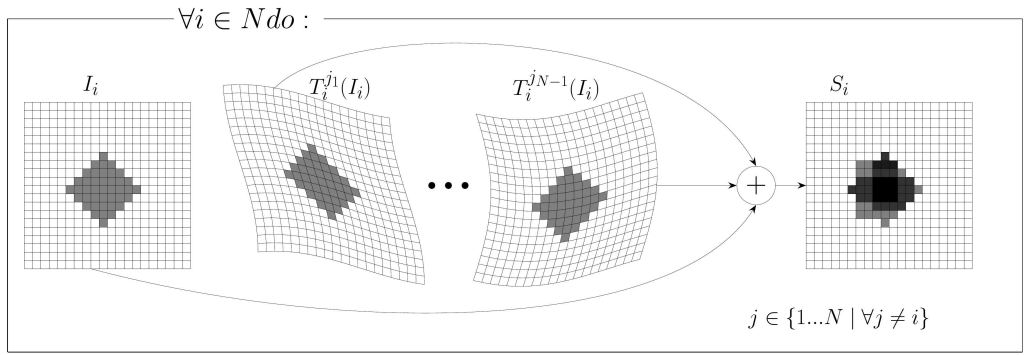}
  \caption{Here the workflow for creating the motion adapted sum of contours $S_i$ is depicted. To create the sum a contour $I_i$ is deformed with the deformation vector field $T^N$ followed by a voxel wise summation of original and deformed contours. $I_i$ is a binary image with 0: background, 1: foreground. The max value of $S_i$ is $N$.}
  \label{fig:sumFlowChart}
\end{figure}

Only the internal datasets contain contours for the average CT, therefore this experiment was restricted to the 11 patients of the internal cohort. 
The current clinical practice to contour OARs on average or ungated CT acquisitions bears the risk that relevant organ motion is either averaged out or not recorded. For this reason, we compared the oesophagus contours delineated on 4DCT, which depict the motion, to the volumes delineated on the average CT images. In a first step we created 11 4D contour sums $S^{N+1}$ for each patient. The first one by simply summing up the contours drawn on each time phase image of the 4DCT. The remaining 10 were 'motion adapted' sums $S^N$ and created for each time phase $I_i \in I^N$. The motion adapted sum $S_i$ for phase $i$ was created by transforming the contour $I_i$ drawn in phase $i$ to all other time phases with the DVFs followed by a summation of the original contour on phase $i$ and the transformed contours. The whole workflow is depicted in Fig. \ref{fig:sumFlowChart} and the pseudo code can be seen in \ref{app:ComparisonofContouringMethods}.

In a second step the coverage of the contours drawn on the 4DCT by the contour drawn on the average CT was measured by calculating DSC, HD and ASSD between the contour delineated on the average CT and all $S_n \in S^{N+1}$. For this we had to normalise $S^{N+1}$ by setting all voxel values above 1 to 1.

\begin{table}[!htb]
\centering
\caption{Overview of the contouring and motion analysis results for the 11 clinical (\textit{C1 - C11}) and the 5 The Cancer Imaging Archive (\textit{T1 - T5}) datasets. The second column gives the ratio between the 4DCT union oesophagus contours and the contour done on average CT. Columns 3 to 5 give Sørensen-Dice index (\textit{DSC}), Hausdorff distance (\textit{HD}) and average symmetric surface distance (\textit{ASSD}). Columns 6 to 8 show the maximum absolute motion left-right (\textit{LR}), anterior-posterior (\textit{AP}) and superior-inferior (\textit{SI}). }
\label{tab:datasetOverview}
\begin{tabular}{lccccccc}
\toprule
\textbf{Data} & {\textbf{$\frac{Vol.\ 4D\  Union}{Vol.\  average} $}} & \textbf{DSC} & {\begin{minipage}{0.5in}\centering\textbf{HD (mm)}\end{minipage}} & \begin{minipage}{0.4in}\centering\textbf{ASSD (mm)}\end{minipage} & \begin{minipage}{0.3in}\centering\textbf{LR (mm)}\end{minipage} & {\begin{minipage}{0.3in}\centering\textbf{AP (mm)}\end{minipage}} & {\begin{minipage}{0.3in}\centering\textbf{SI (mm)}\end{minipage}} \\
\midrule
   C1 & 1.37 & 0.82 & 10.73 & 1.19 & 5.50 & \phantom{0}4.08 & \phantom{0}5.60
  \\
   C2 & 1.41 & 0.79	& 11.15 & 1.30 & 4.44 & \phantom{0}4.77 & 15.98
 \\
   C3 & 1.43 & 0.79	& 12.79 & 1.45 & 4.83 & \phantom{0}4.99 & \phantom{0}5.30
  \\
   C4 & 1.37 & 0.82 & 11.68 & 1.14 & 5.72 & \phantom{0}3.20 & \phantom{0}9.99
 \\
   C5 & 1.22 & 0.84	& \phantom{0}6.83 & 0.89 & 1.16 & \phantom{0}1.08 & \phantom{0}3.47
 \\
   C6 & 1.46 & 0.78	& 12.33 & 1.24 & 3.32 & \phantom{0}4.88 & \phantom{0}9.52
 \\
   C7 & 1.61 & 0.76	& \phantom{0}8.56 & 1.73 & 4.54 & \phantom{0}4.69 & \phantom{0}8.21
 \\
   C8 & 1.41 & 0.73	& 11.05 & 1.40 & 4.70 & \phantom{0}3.66 & \phantom{0}7.87
 \\
   C9 & 1.31 & 0.80	& 13.81 & 1.16 & 2.64 & \phantom{0}2.58 & \phantom{0}4.68
 \\
   C10 & 1.18 & 0.86 & \phantom{0}9.95 & 0.81 & 3.06 & \phantom{0}5.28 & 16.49
 \\
   C11 & 1.42 & 0.81 & \phantom{0}8.78 & 1.24 & 3.33 & \phantom{0}4.48 & \phantom{0}6.64
 \\
   T1 & {-} & {-} & {-} & {-} & 3.79 & \phantom{0}7.72 & 14.34 \\
   T2 & {-} & {-} & {-} & {-} & 4.70 & \phantom{0}9.83 & 17.33 \\
   T3 & {-} & {-} & {-} & {-} & 9.08 & \phantom{0}8.33 & 14.16 \\
   T4 & {-} & {-} & {-} & {-} & 6.75 & 11.13 & 21.38 \\
   T5 & {-} & {-} & {-} & {-} & 5.19 & \phantom{0}3.96 & 11.47 \\
   \midrule
   \textbf{Mean} & \bfseries 1.38 & \bfseries 0.80 & \bfseries 10.70 & \bfseries 1.23 & \bfseries 4.55 & \bfseries \phantom{0}5.29 & \bfseries 10.78
 \\
   \textbf{SD} & \bfseries 0.21 & \bfseries 0.04 & \bfseries \phantom{0}2.88 & \bfseries 0.31 & \bfseries 1.81 & \bfseries \phantom{0}2.67 & \bfseries \phantom{0}5.30
 \\
    \textbf{Min} & \bfseries 1.03 & \bfseries 0.72 & \bfseries \phantom{0}6.22 & \bfseries 0.72 & \bfseries 1.16 & \bfseries \phantom{0}1.08 & \bfseries \phantom{0}3.47
 \\
    \textbf{Max} & \bfseries 1.72 & \bfseries 0.87 & \bfseries 17.43 & \bfseries 1.87 & \bfseries 9.08 & \bfseries 11.13 & \bfseries 21.38
 \\ 
\bottomrule 
\end{tabular}

\end{table}

In a third step we estimated voxel wise the time (in fractions of the respiratory cycle) oesophageal tissue is not covered by the contour of the average CT by setting all voxels in $S_n \in S^{N+1}$ that are covered by average CT contour to zero. A voxel value in such a masked masked sum image indicates in how many time phases oesophageal tissue is present at the coordinates of the respective voxel but not covered by the contour of the average CT.

\begin{figure}[h!]
     \begin{center}
     \mbox{
      \shortstack{
        \includegraphics[height=0.424\linewidth]{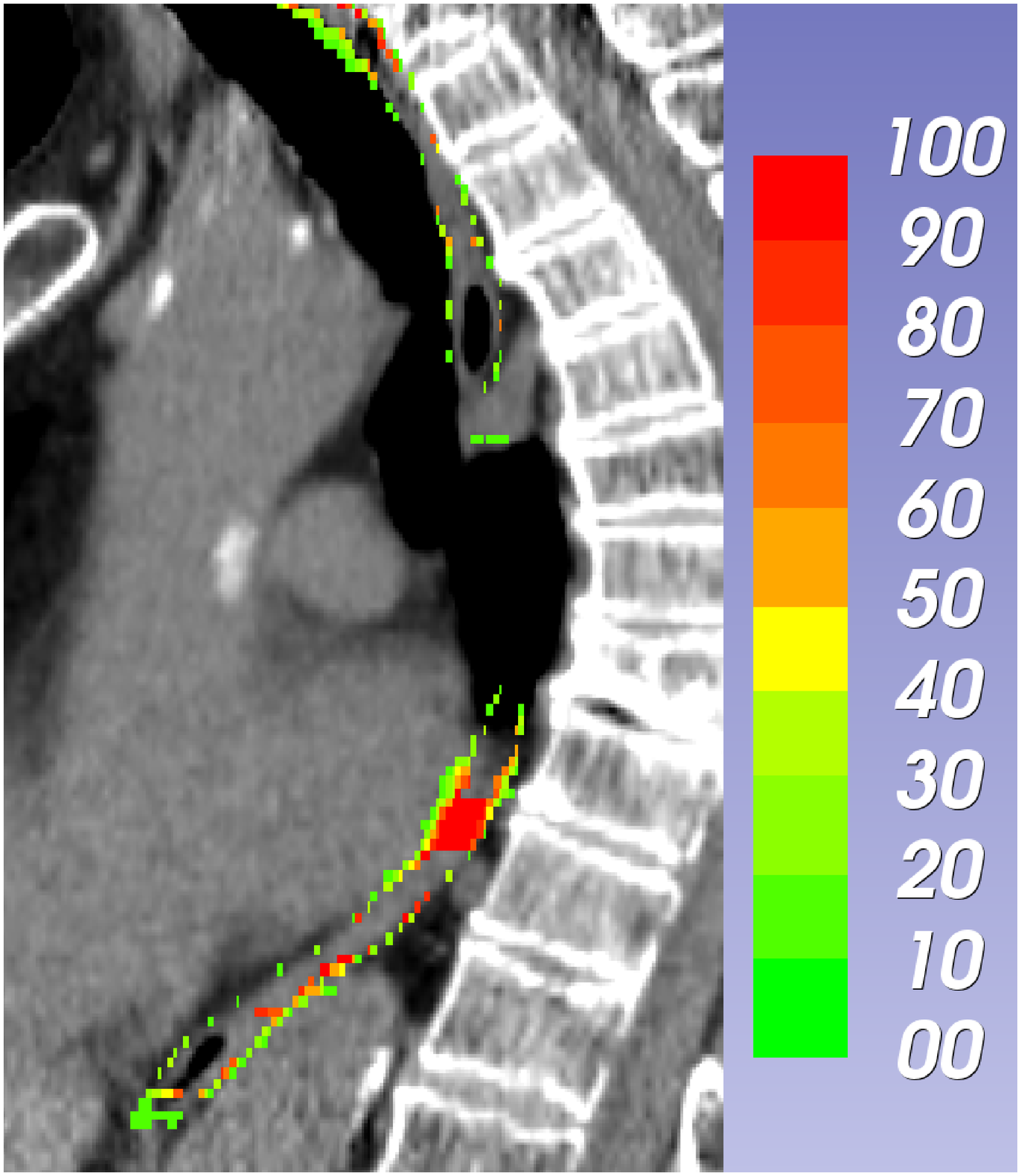} \\
        a) sagittal view of clinical \\ dataset 10 with voxels \\ outside the average CT contour
        }
        \hspace{-1.5 mm}
         
      \shortstack{     
        \includegraphics[width=0.45\linewidth]{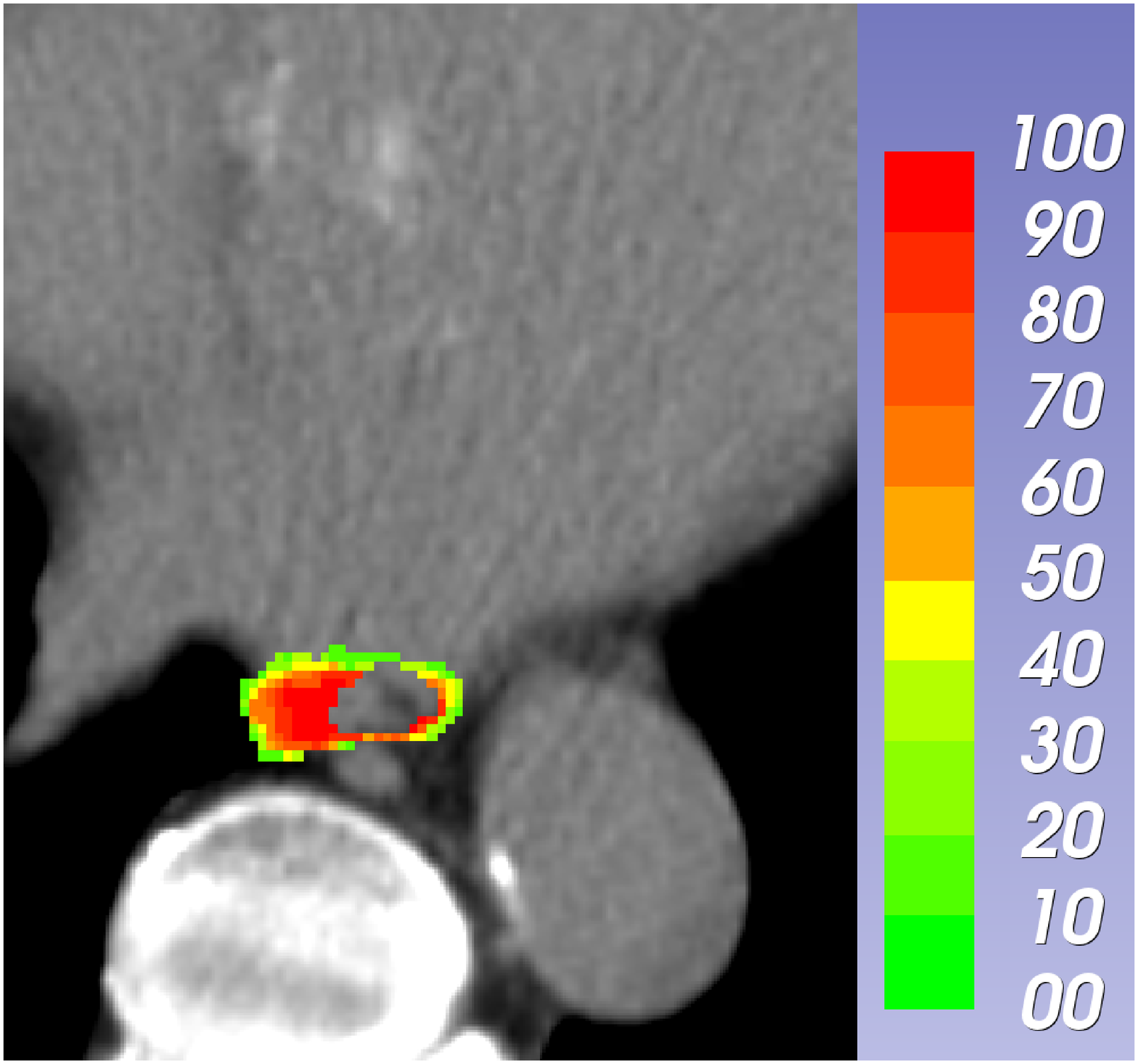} \\
        b) axial view of clinical \\ dataset 10 with voxels \\ outside the average CT contour
        }
               
        }
        
        \mbox{
              \shortstack{     
        \includegraphics[width=0.9\linewidth]{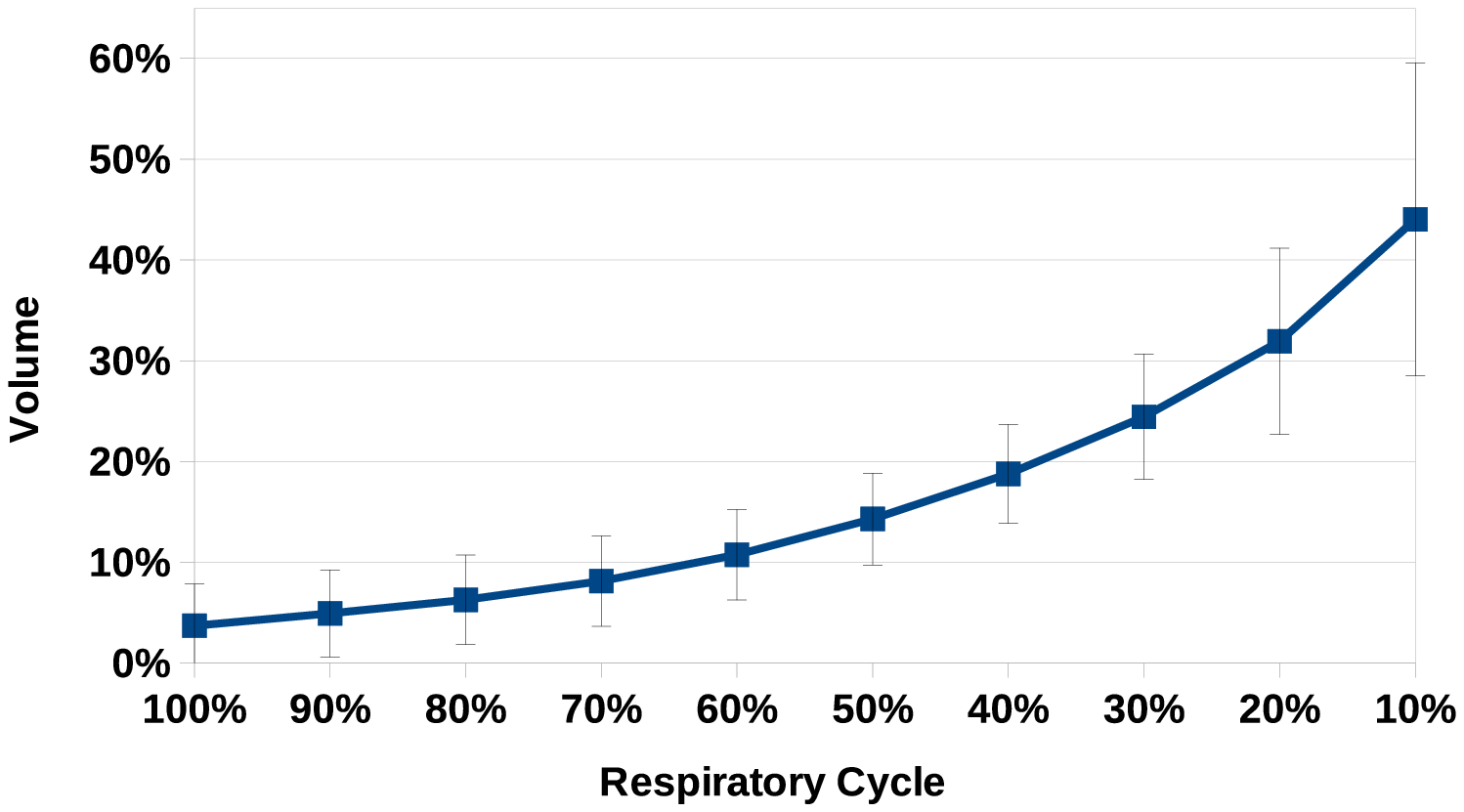} \\
        c) the time (in percent of the respiratory cycle) oesophageal \\ volume (in percent of the average CT contour) resides outside \\ of the average CT contour; vertical bars indicate standard deviation         
        }
               
        }  
    
    \caption{This figure shows oesophageal tissue that was not covered by the contour done on average CT. In a) and b) the color indicates the time (in percent of the respiratory cycle) a voxel was not covered by the average CT contour but depicted oesophageal tissue. c) shows the average over 11 cases for the volume fraction of oesophagus lying outside the contoured volume defined on average CT for a specific fraction of the respiratory cycle.}
\label{fig:VolumeOutsideUngated}                     
\end{center} 
\end{figure}

The comparison between the oesophagus contours on 4DCT and the contour on average CT showed that the volumes of the contour sums $S_i \in S^{n+1}$ are on average 1.38 times bigger than the volume described on average CT. Average DSC, HD and ASSD $\pm$ SD were 0.80 $\pm$ 0.04, 10.70 $\pm$ 2.88 mm and 1.23 $\pm$ 0.31 mm, respectively. Results for each case can be found in Table \ref{tab:datasetOverview}.
The analysis of the masked sum images revealed that around 14 \% of the oesophageal volume was outside the average contour for at least 50 \% of the respiratory cycle. Fig. \ref{fig:VolumeOutsideUngated} illustrates the chronological coverage of the 4D contours by the average contour.

\subsection{Dose Analysis}
This experiment was restricted to the internal cohort because dose information was available only for these patients. As mentioned above, each time phase of the 4DDose contains a fraction of the delivered dose. To facilitate a comparison with the 3DDose of the average CT, we had to map the 4DDose back to a 3DDose volume under consideration of the DVF motion information. For this motion corrected 3DDose volumes $M_i \in M^N$ were created. The voxel value at position $x$ in $M_i$ is calculated by summing up the dose values in the 4DDose $I^N$ along a voxels trajectory given by the DVF $T^N$ starting from $I_i$. Fig. \ref{fig:trajectory} depicts the respiration induced trajectory of a voxel. For every $I_i \in I^N$ one motion corrected dose $M_i$ was created. Further details on the algorithm are given in \ref{app:doseAnal}. 

In the gamma analysis we compared the 3DDose to all motion corrected dose volumes $M_i$. The gamma analysis was done with the contour defined on the average CT. Also the dosimetric indices $D_{max}$, $D_{mean}$, $D_{2\%}$, $D_{33\%}$ and $D_{50\%}$ were calculated for the 3DDose and all $M_i$ but with the contour related to the respective time phase.

\begin{table}[!htb]
\centering
\footnotesize
\caption{Column 2 shows the distance of the oesophagus from the delivered maximum dose for the 11 clinical cases (\textit{C1 - C11}). The average gamma passing rates are given in the last two columns with a 10 \% cut-off of the prescription dose (60 Gy) or the maximum dose inside the oesophagus, respectively.}
\label{tab:doseAnalysesGamma}
\begin{tabular*}{0.75\textwidth}{@{\extracolsep{\fill}}lC{1.0cm}cc}
\toprule
\textbf{Data} & \textbf{Dist (mm)} & \textbf{$3\, \%/2 mm/90 \%$} & \textbf{$3\, \%/3 mm/80 \%$} \\
\midrule
   \rule{0pt}{10pt}C1 & \phantom{0}63  & 0.71 $\pm$ 0.16 & 0.73 $\pm$ 0.07\\
   C2 & \phantom{0}66 & 0.85 $\pm$ 0.12 & 0.81 $\pm$ 0.08\\
   C3 & \phantom{0}65 & 0.50 $\pm$ 0.32 & 0.62 $\pm$ 0.24\\
   C4 & \phantom{0}66 & 0.78 $\pm$ 0.14 & 0.88 $\pm$ 0.04\\
   C5 & \phantom{0}56 & 0.90 $\pm$ 0.08 & 0.85 $\pm$ 0.05\\
   C6 & \phantom{0}58 & 0.84 $\pm$ 0.05 & 0.85 $\pm$ 0.04 \\
   C7 & \phantom{0}79 & 0.78 $\pm$ 0.07 & 0.84 $\pm$ 0.03 \\
   C8 & \phantom{0}65 & 0.93 $\pm$ 0.06 & 0.74 $\pm$ 0.05\\
   C9 & \phantom{0}38 & 0.99 $\pm$ 0.04 & 0.76 $\pm$ 0.06\\
   C10 & \phantom{0}50 & 0.95 $\pm$ 0.04 & 0.86 $\pm$ 0.05\\ 
   C11 & 100 & 0.96 $\pm$ 0.06 & 0.73 $\pm$ 0.05\\
   \midrule
   \rule{0pt}{10pt}\textbf{Mean} & \bfseries \phantom{0}64 & \bfseries 0.84 & \bfseries 0.79 \\ 
   \textbf{SD} & \bfseries \phantom{0}16 &  \bfseries 0.17 & \bfseries 0.13 \\
   \textbf{Max} & \bfseries 100 & \bfseries 0.99 & \bfseries 0.88 \\
   \textbf{Min} & \bfseries \phantom{0}38 & \bfseries 0.50 & \bfseries 0.62 \\
\bottomrule     
\end{tabular*}
\end{table}

\begin{figure}[htb]
     \begin{center}
     \mbox{
      \shortstack{
        \includegraphics[width=0.3\linewidth]{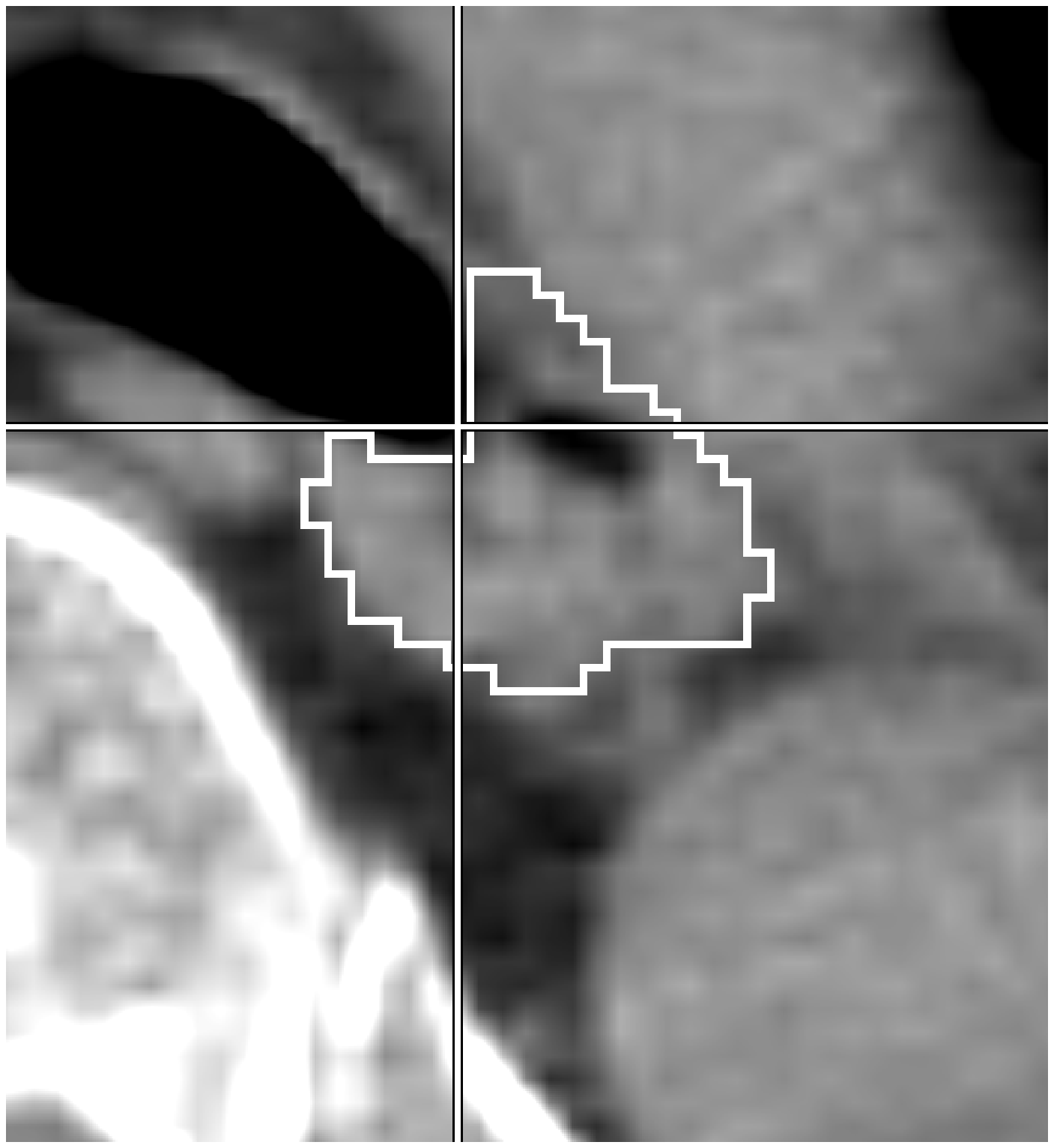} \\
       a) average CT
        }

      \shortstack{     
        \includegraphics[width=0.3\linewidth]{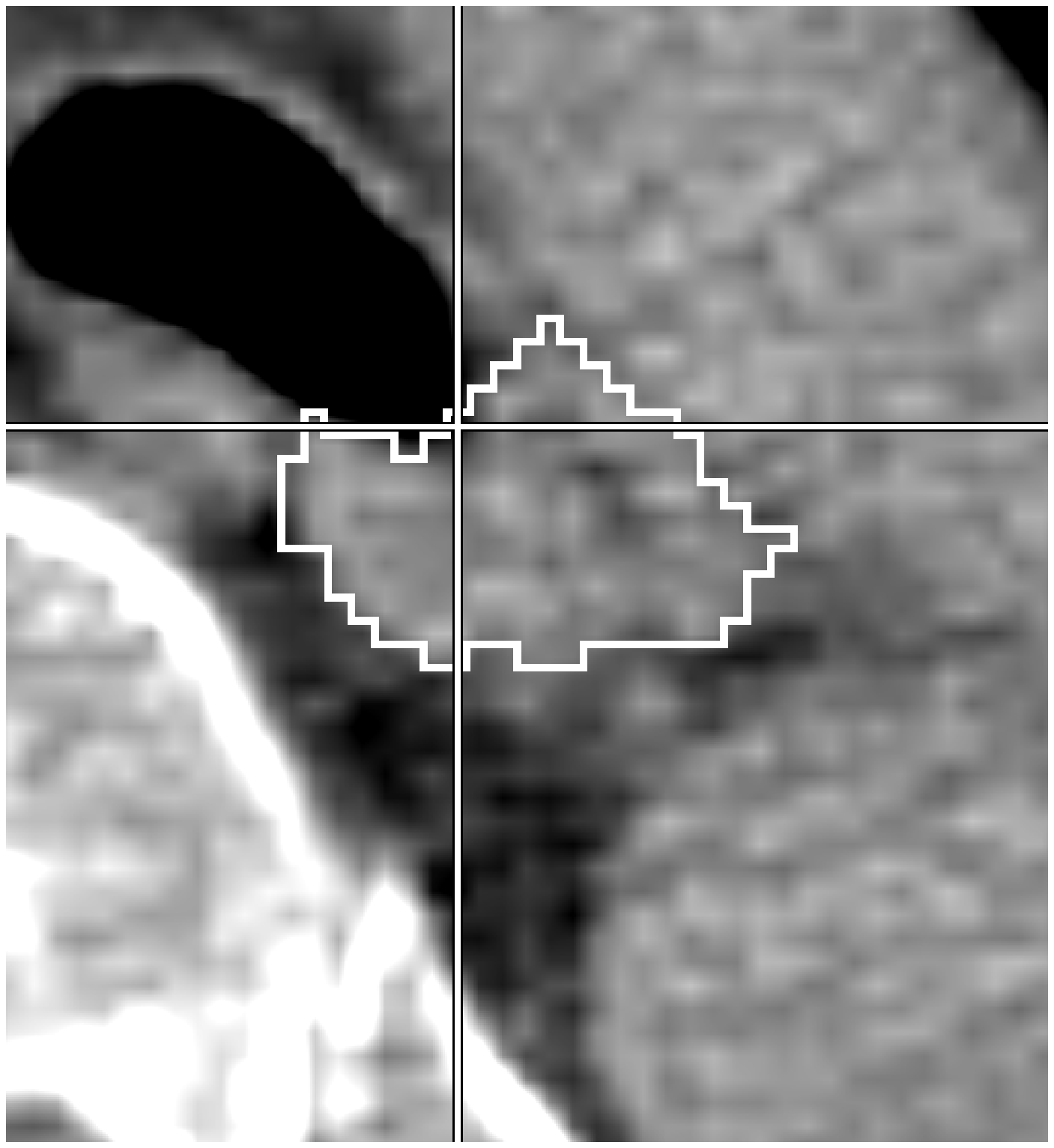} \\
        b) time phase 3
        }
        
      \shortstack{     
        \includegraphics[width=0.3\linewidth]{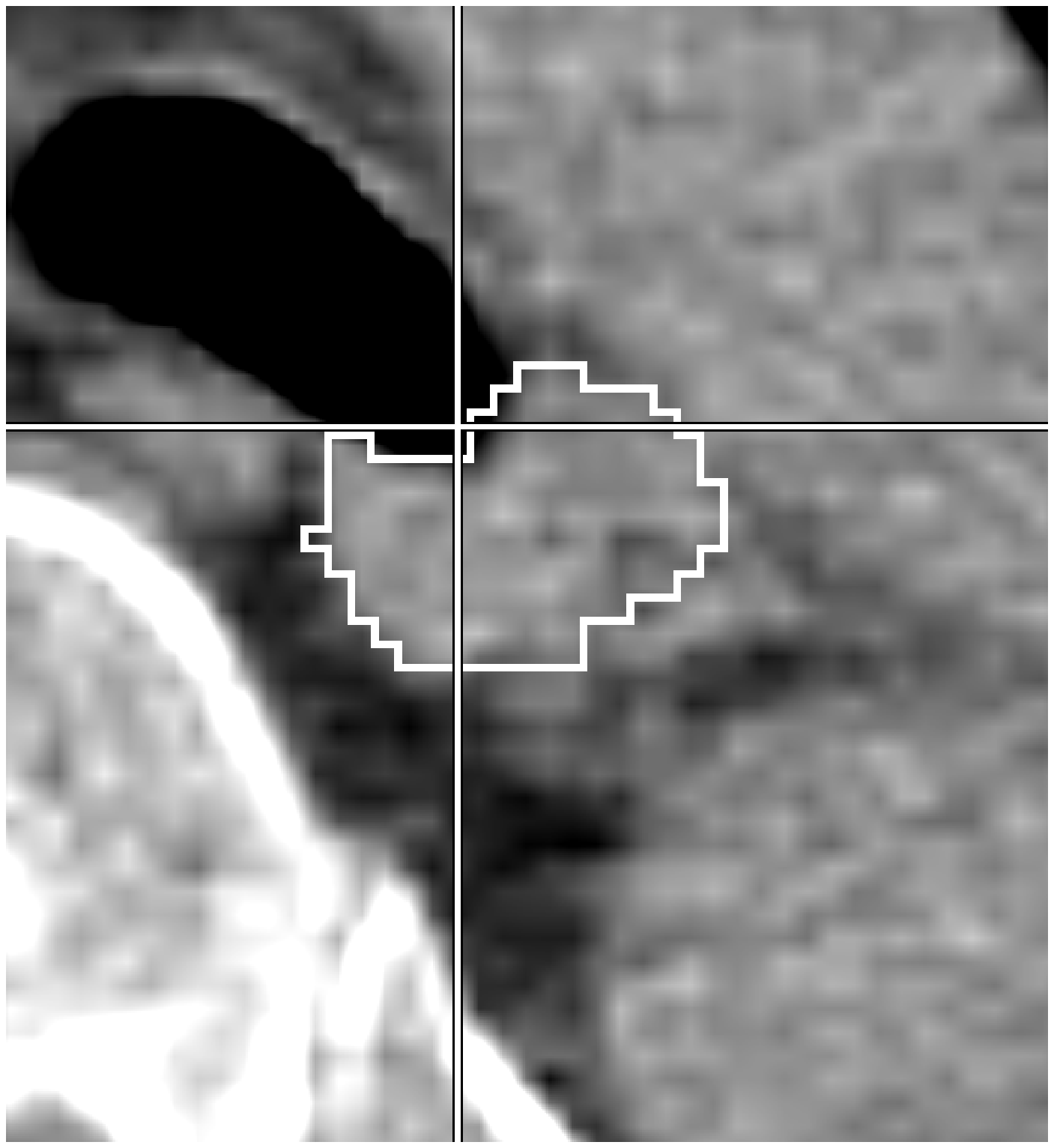} \\
        c) time phase 6
        }        
               
        }

        \caption{An illustration of the inconsistencies between contoured organ and real underlying motion for patient C4. With the use of the cross-hair it can be seen that the tissues moves from time phase 3 b) to time phase 6 c) towards the lower right image corner, but the shrinkage of the contour from time phase 3 to time phase 6 could be interpreted as an indicator for the opposite direction.}
\label{fig:contoursC4}
\end{center}        
\end{figure}

The investigation of the impact of motion on delivered dose resulted in statistically significant different dose distributions of average CT and motion corrected dose for all cases. However, the more clinically relevant gamma comparisons (Table \ref{tab:doseAnalysesGamma}) showed a violation of the gamma criterion in 6 out of 11 cases for $3\, \%/2 mm/90 \%$ and in 5 out of 11 cases for $3\, \%/3 mm/80 \%$. The maximal measured deviation of the maximum dose was 2 Gy, whereas $D_{2\%}$ varied maximally by 1 Gy. In 9 of the 11 patients $D_{max}$ calculated with the average CT was covered by the 95 \% confidence interval of the motion corrected $D_{max}$ values. For $D_{2\%}$ this was the case for 10 of the 11 patients. The dosimetric values are summarized in Table \ref{tab:doseAnalyses}. Differences between the two gamma experiments (e.g in C9 and C11) occurred because the cut-off threshold calculated on basis of the prescription dose was close to the maximum dose inside the organ and therefore solely a small amount of voxels could be considered for the calculations. For some cases, e.g. C4, the motion corrected dose showed lower and higher dose values compared to the dose distribution calculated on average CT. We can attribute this phenomenon to uncertainties in the organ delineation. In Fig. \ref{fig:contoursC4} it is shown that the tissue motion is not reflected by the contours and by considering just the contours, one could assume motion in a different direction.  

\begin{landscape}
\begin{table}[htb]
\footnotesize
\caption{Dosimetric analysis for the 11 clinical (\textit{C1 - C11}) datasets. \textit{$D_{max}$ Average} lists the maximum dose values the oesophagus received according to the calculation on the average CT scan. The motion corrected dose was calculated for each time phase contour, the range of the maxima and mean values are given in \textit{$D_{max}$} and \textit{$D_{mean}$ Motion}. $D_{x\%}$ ($x \in \{2, 33,50\}$) give the dose received by at least $x \%$ of the volume.}
\label{tab:doseAnalyses}
\begin{tabular}{L{1.0cm}C{1.0cm}C{1.8cm}C{1.0cm}C{1.8cm}C{1.0cm}C{1.8cm}C{1.0cm}C{1.8cm}C{1.0cm}C{1.8cm}}
\toprule
\textbf{Data} 
& \multicolumn{2}{c}{{\textbf{$D_{max}$ (Gy)}}}
& \multicolumn{2}{c}{{\textbf{$D_{mean}$ (Gy)}}} 
& \multicolumn{2}{c}{{\textbf{$D_{2\%}$ (Gy)}}}
& \multicolumn{2}{c}{{\textbf{$D_{33\%}$ (Gy)}}} 
& \multicolumn{2}{c}{{\textbf{$D_{50\%}$ (Gy)}}} 
\\
& \textbf{Average} & \textbf{Motion}
& \textbf{Average} & \textbf{Motion} 
& \textbf{Average} & \textbf{Motion}
& \textbf{Average} & \textbf{Motion} 
& \textbf{Average} & \textbf{Motion} 
\\
\midrule
   \rule{0pt}{10pt}C1 &  16.44 & 16.08-16.67 & 2.60 & 2.45-2.84 & 13.84 & 13.47-13.85 & \phantom{0}1.05 & 1.04-1.25 & 0.58 & 0.59-0.64\\
   C2 & 10.36 & 9.86-10.73 & 2.32 & 2.13-2.59 & \phantom{0}9.09 & 8.82-9.15 & \phantom{0}1.57 & 1.17-3.05 & 0.55 & 0.50-0.69\\
   C3 & 11.45 & 11.01-11.41 & 0.96 & 0.82-1.16 & \phantom{0}9.21 & 9.16-9.74 & \phantom{0}0.28 & 0.25-0.29 & 0.15 & 0.14-0.17\\
   C4 & 37.70 & 35.69-39.54 & 8.59 & 7.52-8.58 & 30.17 & 28.33-30.31 & 13.90 & 10.29-13.81 & 1.36 & 1.53-1.92\\
   C5 & 16.03 & 16.04-16.55 & 4.24 & 4.15-4.47 & 14.62 & 14.59-14.88 & \phantom{0}5.57 & 4.67-6.78 & 0.68 & 0.65-0.71\\
   C6 & 24.57 & 22.68-25.02 & 4.64 & 4.02-4.93 & 17.88 & 17.12-18.65 & \phantom{0}3.15 & 2.37-3.60 & 1.08 & 0.57-1.18\\
   C7 & 18.71 & 17.86-19.69 & 4.42 & 4.28-4.70 & 15.29 & 14.90-15.87 & \phantom{0}6.75 & 6.19-7.72 & 0.94 & 0.87-1.09\\
   C8 &  \phantom{0}9.18 & 9.86-10.56 & 1.27 & 1.39-1.68 & \phantom{0}8.08 & 8.65-9.08 & \phantom{0}0.42 & 0.44-0.55 & 0.21 & 0.18-0.28\\
   C9 & \phantom{0}7.14 & 5.84-7.19 & 1.11 & 0.74-1.26 & \phantom{0}6.06 & 4.99-5.99 & \phantom{0}0.37 & 0.30-0.43 & 0.23 & 0.20-0.24\\
   C10 & 10.96 & 10.42-11.32 & 3.23 & 2.92-3.40 & \phantom{0}9.62 & 9.45-9.70 & \phantom{0}5.90 & 4.04-6.44 & 0.71 & 0.51-0.85\\ 
   C11 & \phantom{0}8.49 & 8.15-8.42 & 1.30 & 1.06-1.30 & \phantom{0}7.50 & 7.32-7.49 & \phantom{0}0.40 & 0.35-0.40 & 0.24 & 0.21-0.25\\
   \midrule
   \rule{0pt}{10pt}\textbf{Mean} & \bfseries 15.55 & \bfseries 15.42 & \bfseries 3.15 & \bfseries 3.10 & \bfseries 12.85 &  \bfseries 12.80 & \bfseries \phantom{0}3.58 &  \bfseries \phantom{0}3.37 & \bfseries 0.61 & \bfseries 0.63\\ 
   \textbf{SD} & \bfseries \phantom{0}9.00 & \bfseries \phantom{0}8.57 & \bfseries 2.27 & \bfseries 2.06 & \bfseries \phantom{0}6.86 &  \bfseries \phantom{0}6.45 & \bfseries \phantom{0}4.22 &  \bfseries \phantom{0}3.58 & \bfseries 0.40 &  \bfseries 0.44\\
   \textbf{Max} & \bfseries 37.70 & \bfseries 39.54 & \bfseries 8.59 & \bfseries 8.58 & \bfseries 30.17 &  \bfseries 30.31 & \bfseries 13.90 & \bfseries 13.81 & \bfseries 1.36 & \bfseries 1.92\\
   \textbf{Min} & \bfseries \phantom{0}7.14 & \bfseries \phantom{0}5.84 & \bfseries 0.96 & \bfseries 0.74 & \bfseries \phantom{0}6.06 &  \bfseries \phantom{0}4.99 & \bfseries \phantom{0}0.28 &  \bfseries \phantom{0}0.25 & \bfseries 0.15 & \bfseries 0.14\\
\bottomrule
      
\end{tabular}
 
\end{table}
\end{landscape}

\subsection{Oesophagus Motion Analysis}
\label{subsec:motionAnal}
\begin{figure}[htb]
     \begin{center}
     \mbox{
      \shortstack{
        \includegraphics[width=0.3\linewidth]{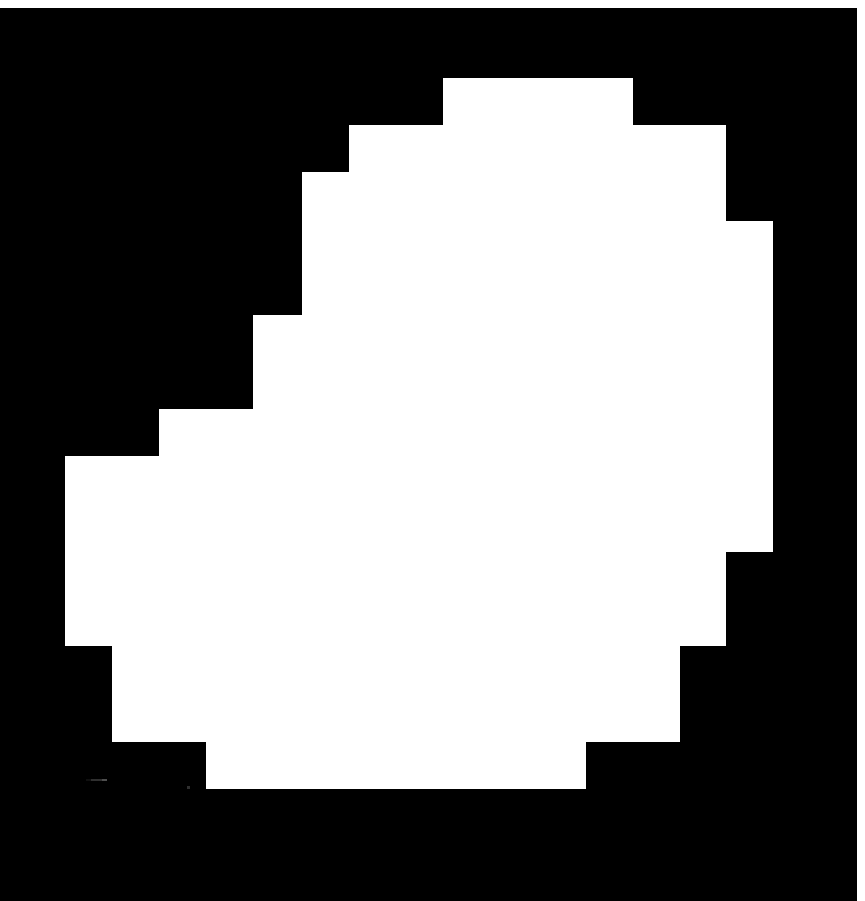} \\
       a) original contour
        }

      \shortstack{     
        \includegraphics[width=0.3\linewidth]{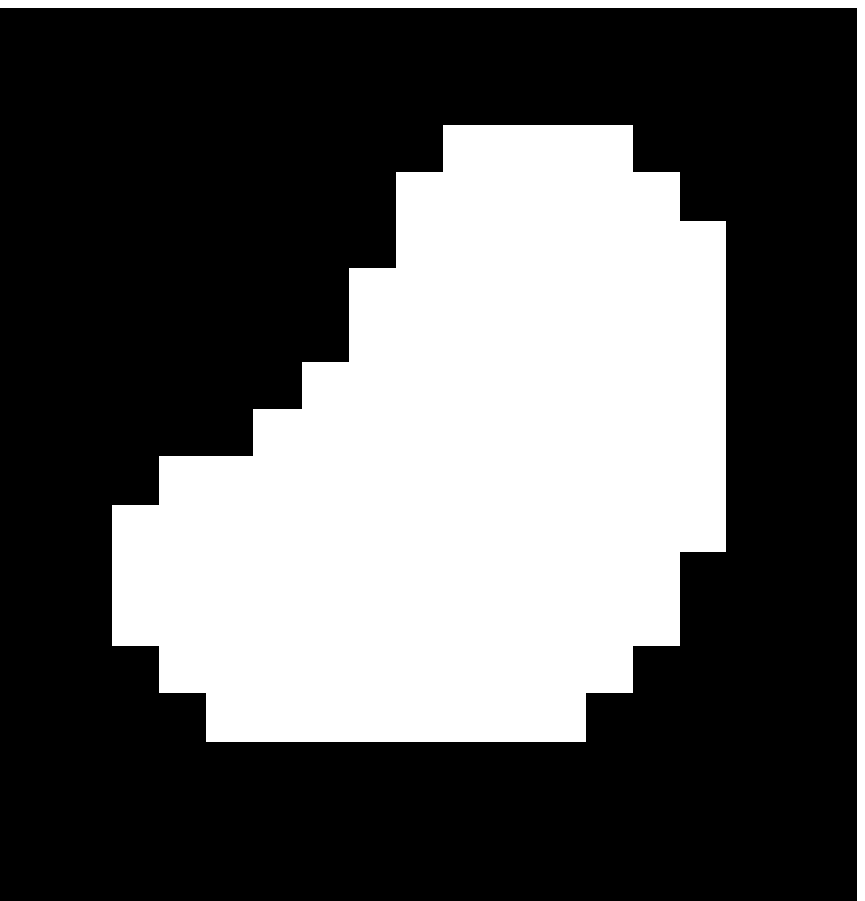} \\
        b) eroded contour
        }
        
      \shortstack{     
        \includegraphics[width=0.3\linewidth]{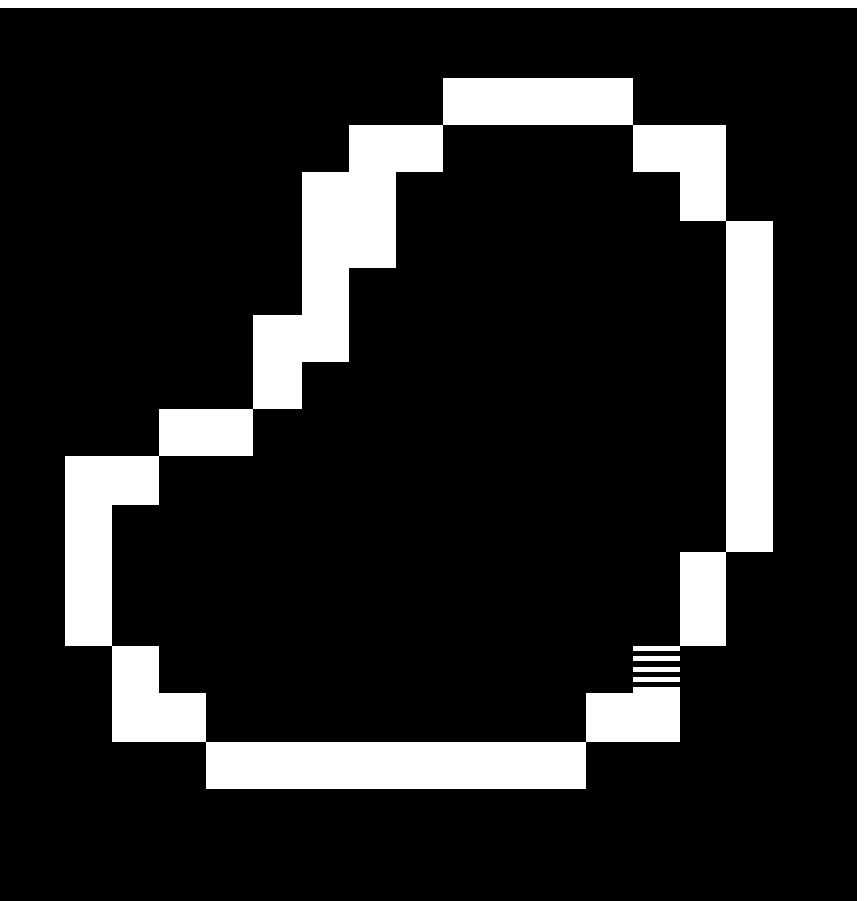} \\
        c) border region
        }        
               
        }

        \caption{In a) an axial view of an oesophagus contour in binary image representation (black: 0, white: 1) can be seen. In b) the eroded contour is depicted and c) shows the border region gained by substracting the eroded contour from the original contour. The small striped area in the lower right of c) indicates the size of one voxel.}
\label{fig:binaryImg}
\end{center}        
\end{figure}

During image acquisition the oesophagus can change its inner appearance e.g. because the patient swallows regions filled with air can appear or disappear. Such changes in the inner appearance are not necessarily connected with a change in the outer shape or position of the organ but are reflected by high amplitudes in the DVFs. To avoid that these high amplitudes which are not caused by a real positional change affect the motion statistics, only the oesophageal border region was used for motion analysis. The border region was created by subtracting the eroded contour from the original contour (Fig. \ref{fig:binaryImg}).

\begin{figure}[!htbp] 
  \centering
     \includegraphics[width=\textwidth]{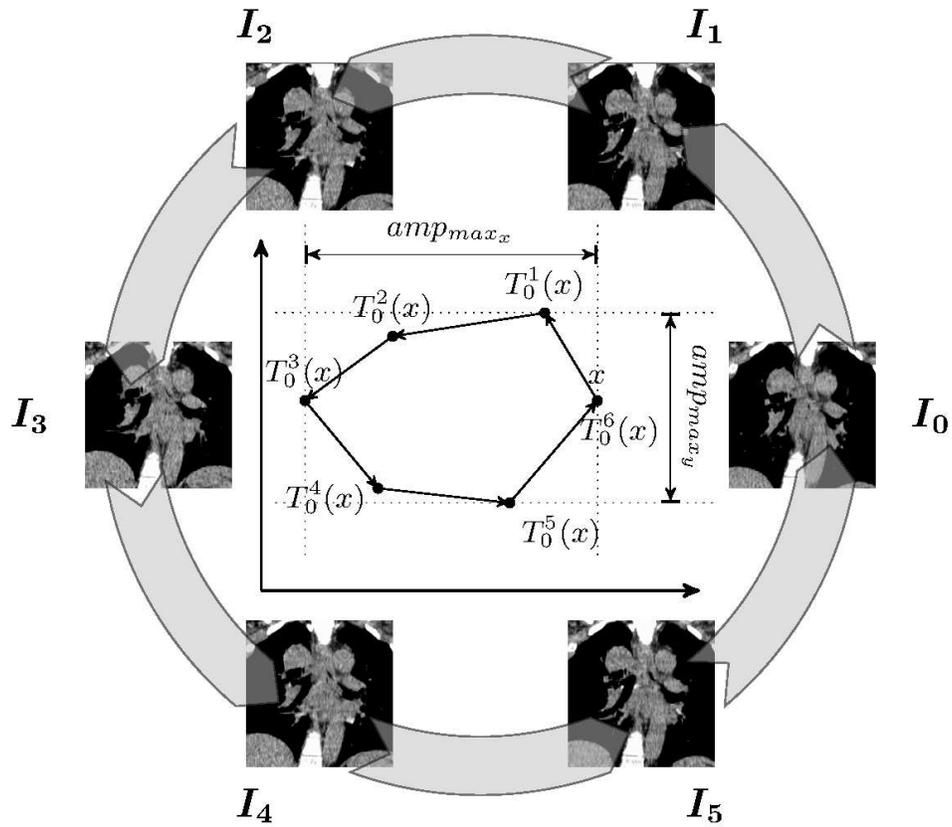}
  \caption{The pathway of a voxel during the respiratory cycle is depicted in the center of the figure. For simplicity the breathing cycle is divided into 5 phases in contrast to the 10 phases of the used datasets. $amp_{max}$ gives the maximal motion amplitude in all dimensions. }
  \label{fig:trajectory}
\end{figure}

The maximal motion extent $amp_{max}$ for a voxel $x$ in the oesophageal border region defined in time phase $I_i$ was estimated by following $x$ on its trajectory in $T^N$ and tracking its position. An illustration how $amp_{max}$ is calculated is given in Fig. \ref{fig:trajectory}. The pseudo code can be found in \ref{app:motionAnal}. $amp_{max}$ was calculated separately in x-, y- and z-direction for all border region voxels and all $I_i \in I^N$.

On average the maximal absolute change in position of the oesophageal border region was 4.55 $\pm$ 1.81 mm, 5.29 $\pm$ 2.67 mm and 10.78 $\pm$ 5.30 mm in LR, AP and SI direction, respectively (see Table \ref{tab:datasetOverview}). What attracts attention is that there is a statistically significant difference between the two patient groups. For example, the TCIA cohort showed a mean maximal absolute motion (\textit{MMAM}) of 8.19 $\pm$ 2.72 mm AP, whereas the internal patients exhibit a value of 3.97 $\pm$ 1.26 mm. We were able to measure a similar but less salient pattern in the other directions: The LR and SI MMAM were 3.93 $\pm$ 1.36 mm and 8.52 $\pm$ 4.31 mm for our internal patients compared to a MMAM of 5.90 $\pm$ 2.08 mm, 15.73 $\pm$ 3.78 mm for the TCIA patients. 

In 7 out of 16 patients the maximum SI motion was measured in the lower part of the oesophagus in the area of the diaphragm. For the remaining 9 patients the SI maximum could be located either directly in the vicinity of the Carina or between Carina and diaphragm. Two LR maxima were located above the Carina, 5 at the height of the Carina, 5 between Carina and diaphragm and 4 in the lower part of the organ in the area of the diaphragm. In AP the location of the maximum positions was similar. In 3 cases the maximum was located above the Carina, 3 cases showed the maximum at the height of the Carina, 3 between Carina and diaphragm and 7 in the area of the diaphragm. An overview of motion amplitudes in connection with maxima positions is given by our motion model in \ref{subsec:motionModel}.

\subsection{Oesophagus Motion Model}
\label{subsec:motionModel}
In recent works the inter-patient oesophageal motion comparison was facilitated by means of comparing the average motion in predefined anatomical regions\cite{cohen2010,Sekii2018,doi2018} or only in the region around the tumor \cite{Yaremko2008,Wang2019}. In this work we envisaged a method for a more detailed comparison on a voxel level. For this purpose we registered the maximum inhale phase of all datasets to a common reference. As reference case we chose Case 100 of TCIA because of its good image quality and it shows the whole oesophagus without any anatomical abnormalities.
The affine registration matrix and the DVFs were then used to transfer the motion information of each patient to the reference case. This procedure allowed for a statistical motion analysis of the whole cohort on a voxel level and a visualization of the whole organ motion easy to interpret.

\begin{figure}[!htb]
     \begin{center}
     \mbox{
      \shortstack{
        \includegraphics[width=0.24\linewidth]{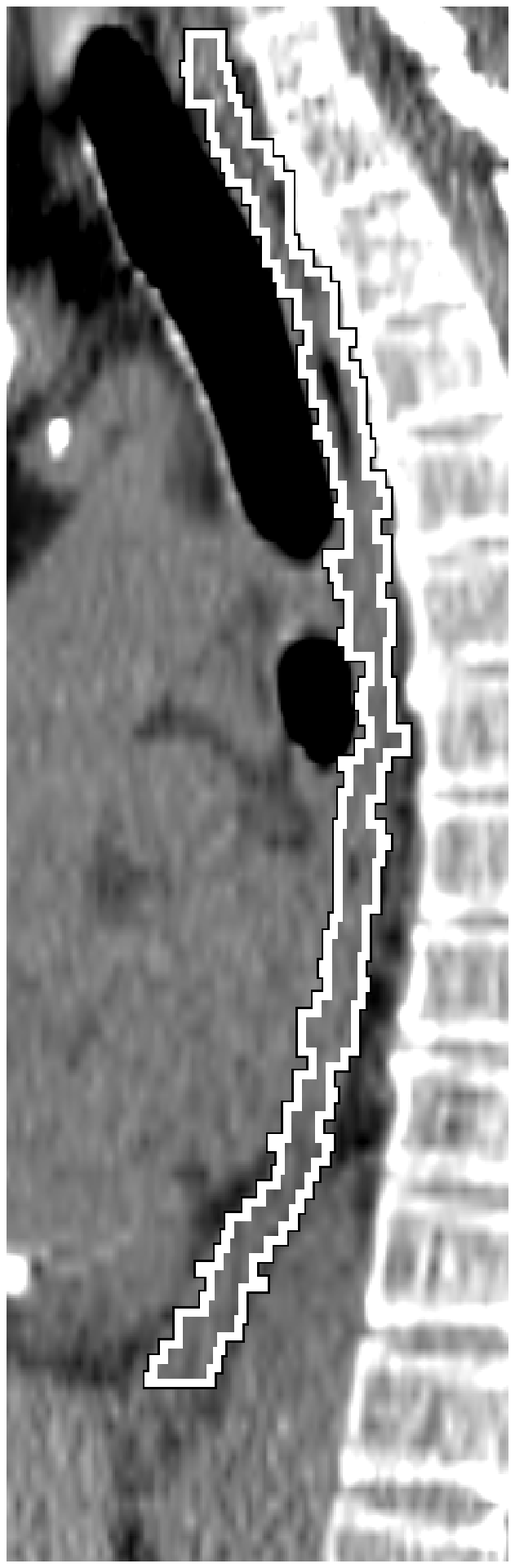} \\
       a) Reference Case
        }
             
      \shortstack{     
        \includegraphics[width=0.24\linewidth]{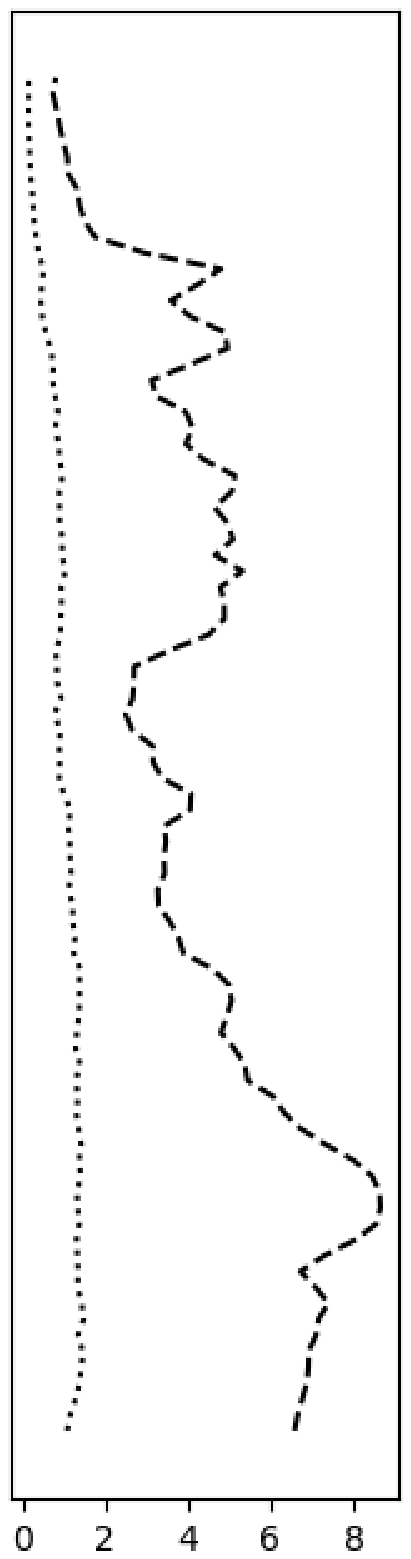} \\
        b) LR Motion (mm)
        }
        
      \shortstack{     
        \includegraphics[width=0.24\linewidth]{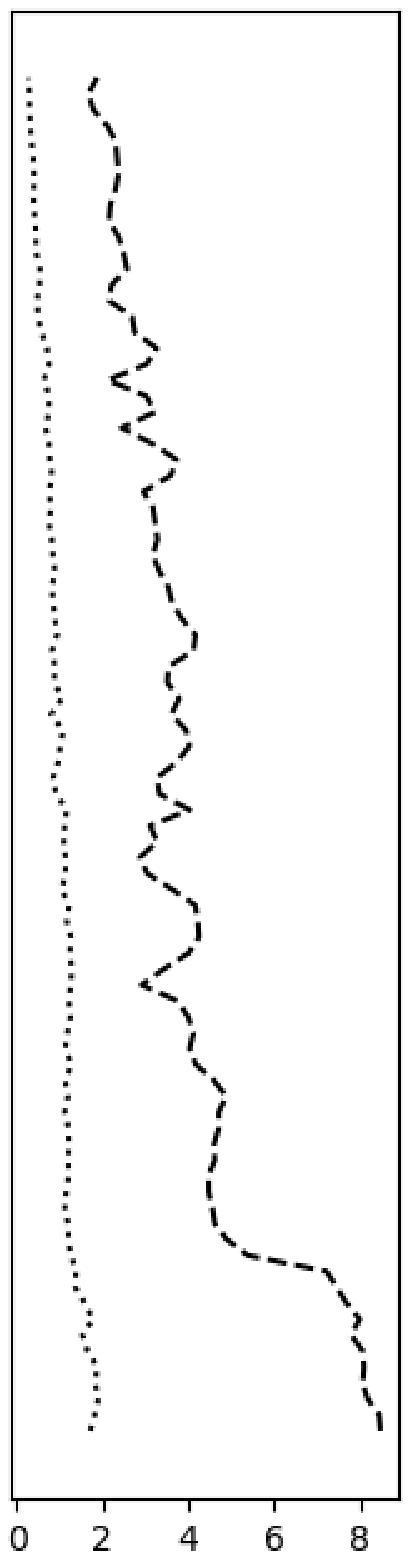} \\
        c) AP Motion (mm)
        }  
        
      \shortstack{     
        \includegraphics[width=0.24\linewidth]{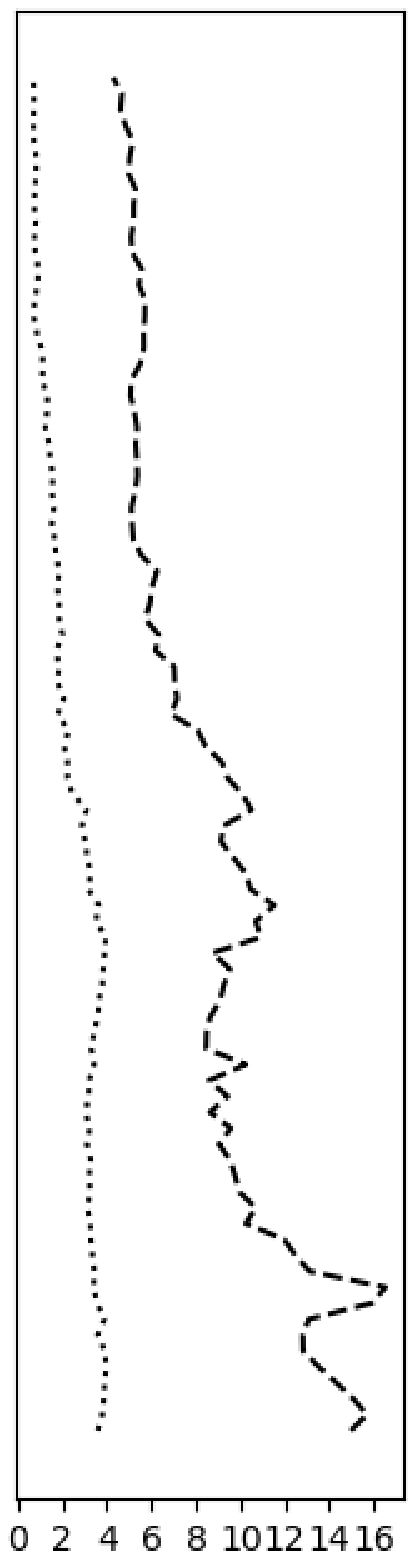} \\
        d) SI Motion (mm)
        }                  
               
        }

        \caption{In a) the sagittal view of the reference case T1 with delineated oesophagus is shown. b) to d) illustrate maximum (dashed) and average (dotted) motion of our model in LR, AP and SI direction at the respective height of the organ. The model was generated by mapping the motion vector fields of all patients to the reference case}
\label{fig:motionModel}
\end{center}        
\end{figure}

\begin{figure}[!htb]
     \begin{center}
     \mbox{
      \shortstack{
        \includegraphics[width=0.44\linewidth]{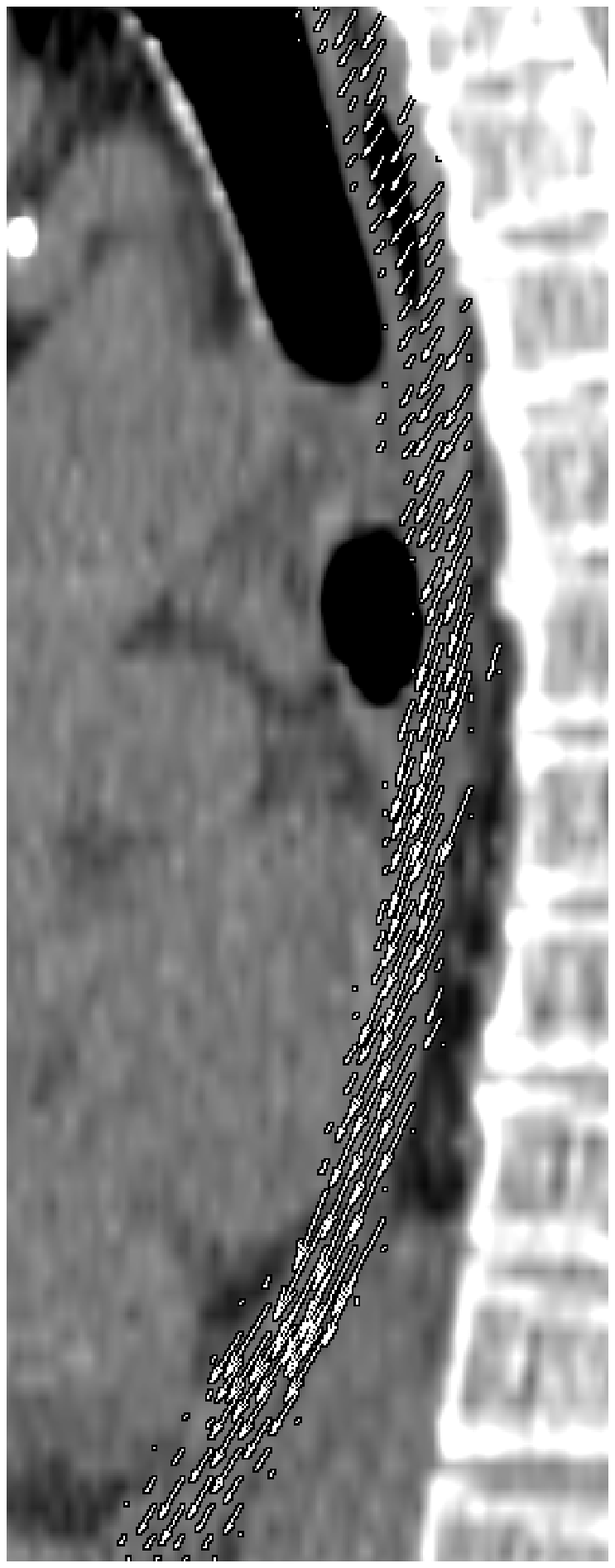} \\
       a) Maximal Motion Sagittal
        }
             
      \shortstack{
              \includegraphics[width=0.44\linewidth]{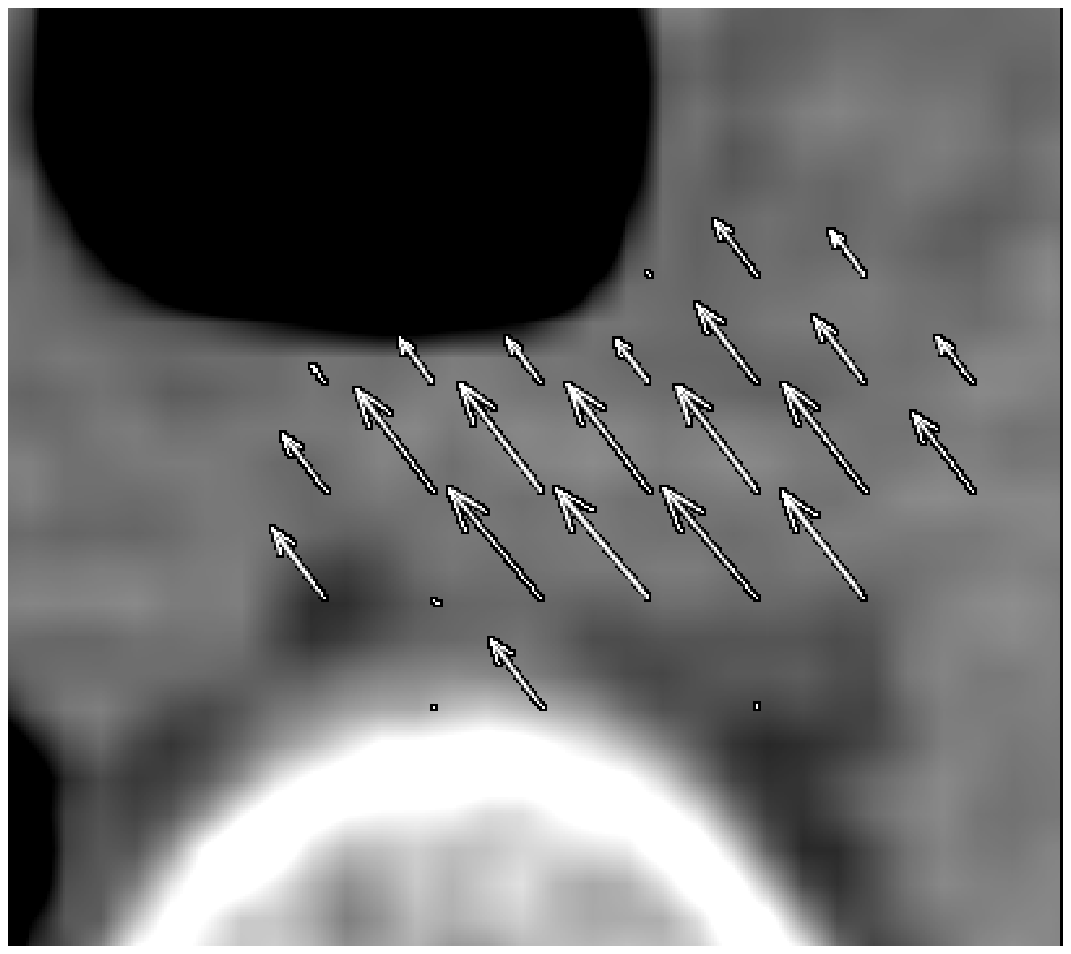}\\
        b) Maximal Motion Axial\\
      	\includegraphics[width=0.44\linewidth]{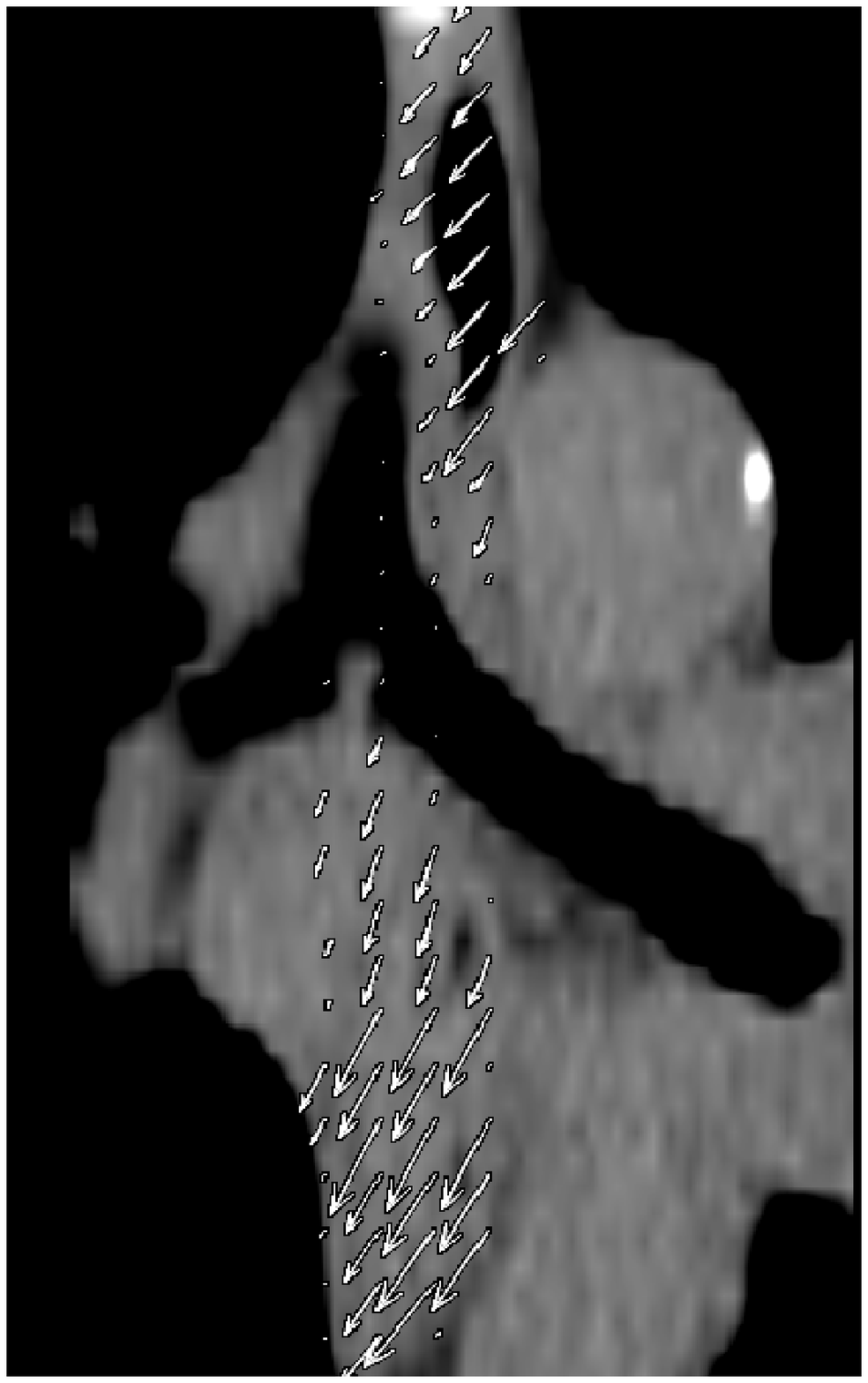}\\
        c) Maximal Motion Coronal
        }

        }

        \caption{The reference case T1 of the created motion model shown in sagittal a), axial b) and coronal c) view. The overlaid arrows indicate the maximal motion of the model. The motion model contains the oesophageal motion information of all 16 patients.}
\label{fig:motionModelDVF}
\end{center}        
\end{figure}

Fig. \ref{fig:motionModel} and \ref{fig:motionModelDVF} show the calculated motion model. The model contained the motion information from all 16 patients. The biggest movement could be expected in the lowest part of the organ. The LR motion exhibited also a local maxima in the upper part of the organ around the Carina (Fig. \ref{fig:motionModelDVF}) which is in concordance with the findings of the \textit{oesophagus motion analysis} experiments. The average motion in LR and AP direction resided around 2 mm, for SI movement the average lay between 2 and 4 mm. However, the maximum motion could be much higher: more than 8 mm in LR and AP direction and more than 16 mm in SI direction. The fact that the patient-wise measured maxima were above the maxima of the model was because of two reasons: First, for the model we used the motion information from one time-phase contour, a maxima that was e.g. measured with the contour of a different time phase is therefore not reflected in the model. Second, the motion analysis was conducted only in the border region of the oesophagus and deviations in the DVF might map the maxima to a voxel outside the border region on the reference case. However, we analysed the DVFs for motion measurement before and after mapping to the reference case and could see only slight changes in the motion distribution (the maxima changed by 0.39 mm, 0.36 mm and 1.28 mm in LR, AP and SI direction, respectively).

\section{Discussion}

In this work we measured oesophageal motion voxel-wise on 4DCT datasets whilst taking into account the periodic breathing motion pattern. Thus we were able to analyse the organ motion for 16 patients and inspect possible under or over dosage of the oesophagus due to its movement. Additionally we scrutinised the delineation of the oesophagus on average CT with a comparison to contouring on 4DCT. In a final step we showed how to create a motion model that facilitates an easier analysis, illustration and comparison of oesophageal motion across patients on a voxel level. 

The results of the contouring methods showed that the contour on average CT does not cover the whole organ and that a big part of the oesophagus resides outside the contour created on the average CT for a significant amount of time. This bears the risk of delivering too much dose to the oesophagus during treatment which could harm especially the radio-sensitive mucosa. The quantification of this motion showed that breathing induced LR MMAM, which is especially of interest for SBRT of the lungs, is on average close to 5.00 mm. This value is much higher than in other publications \cite{Kobayashi2016,Gao2019,Sekii2018,Yaremko2008,doi2018}, solely \citet{cohen2010} reported a similar value. Whereas \citet{cohen2010} reported a lower value (2.80 mm) AP compared to the 5.29 mm of our experiments. A possible explanation of the different values could be that we used DIR and examined the whole organ whereas most other studies used rigid registration, centroid measurements or investigated only a smaller part of the organ. Another point is that the patient cohorts in the literature encompass mostly oesophageal cancer patients, where the tumour could reduce the mobility of the oesophagus. The work by \citet{WEISS200844} examined lung cancer patients and reported an average motion amplitude in axial plane comparable to the one of our internal cohort. Interestingly we experienced a gap in motion amplitudes between our two patient cohorts. An explanation for the different motion amplitudes in our internal and the TCIA cohort could be that the TCIA cohort consists mainly of locally advanced NSCLC with mediastinum infiltration, which is not the case for our internal cohort. Why oesophageal cancer patients show little motion compared to NSCLC patients with infiltrated mediastinum can only be speculated about: maybe larger lymph-nodes or tumour infiltration of locally advanced NSCLC in the mediastinum lead to displacement effects, that increase the mobility of the oesophagus during the breathing cycle while infiltrating oesophagus tumours might be more likely to fix the organ to the mediastinum. Summarized, our motion and contouring analysis suggests that oesophageal motion is heterogeneous and should be measured individually by using timely resolved imaging.

The dose analysis showed that the absolute dose differences between the dose calculated on average CT and the motion corrected dose were all below or equal 2 Gy. At this point it is important to note that the mean distances of the oesophagus to the dose maximum was 64 mm and the average maximum dose inside the oesophagi around 15 Gy. A treatment of tumours closer to the oesophagus and higher dose gradients in the vicinity of the organ would probably increase the difference between planned and motion corrected dose. In approximately $50\, \%$ of the cases the gamma passing-rate criterion was not fulfilled which could be an indication for severe deviations from the planned dose. Additionally, for almost one fifth of the patients the maximum dose calculated on the average CT was not within the $95\, \%$ confidence interval of the motion corrected maximum dose. Although the results are connected to uncertainties from e.g. segmentation, registration or dose calculations, the trends seen in the results in combination with the actions we took to reduce the influence of those uncertainties (as discussed later on) might indicate that oesophageal motion should be considered also in the planning process.

With the created motion model, which encompasses the information from all patients we are able to illustrate the spatial distribution of the motion precisely for our two cohorts. The model shows that for some cases the maximal motion amplitude (especially in the axial plane) can be located around the Carina. Other studies split the oesophagus in different areas for motion analysis. Often the Carina was used as a boundary for those areas \cite{cohen2010,Sekii2018,doi2018}, maxima located above and below the Carina or in its direct vicinity could lead to averaging out and underestimating the motion.

Factors of uncertainty in our experiments are the created oesophagus demarcations and the image registration matrices and DVFs. Fig. \ref{fig:contoursC4} shows that although we double checked every contour, inconsistencies due to misinterpretations can never be fully excluded. We tackled this issue by repeating the experiments for motion and dose inspection for each time phase image separately. The influence of registration or contouring mistakes is lowered by averaging over all time phase images. Also the extra QA of our registration algorithm and the comparison of the motion model amplitudes with the motion amplitudes of the original datasets are actions we took to lower uncertainty and improve the quality of our results.

We are aware of the relatively small number of patients enrolled in this study, and results need to be seen cautiously. By further improving our algorithms we are confident that we will be able to reduce the high amount of work necessary for the presented analysis, and to extend our study by more patients, including such with tumours in proximity to the oesophagus, where dose differences due to oesophageal movements might be more relevant, in the near future.

\section{Conclusion}

In this work we investigated the breathing induced motion of the oesophagus for central NSCLC patients. We compared the current clinical practice of contouring the oesophagus on 3DCT to timely resolved delineations and measured the impact of the motion on the delivered dose. Additionally, we build a motion model, encompassing motion information from all patients, for a voxel-wise illustration of the motion along the whole organ. Our results showed that 4D information should be considered in the contouring process and that motion influences the delivered dose. The experiments further showed that high motion amplitudes are not limited to the regions around the diaphragm, vary between patients and should be measured individually for each patient. Summarised, the numbers and tools presented in this work can mitigate motion related uncertainties, which might facilitate, in combination with additional experiments that further prove our initial results, SBRT closer to the mediastinal region.

\section*{Acknowledgements}
The authors appreciate the valuable and constructive input from the editors and reviewers of the journal.

\paragraph*{Disclosure of Conflicts of Interest:} The authors have no relevant conflicts of interest to disclose.

\bibliographystyle{elsarticle-num-names}
\bibliography{paperbibRevision}

\appendix
\section{}
\label{app:ComparisonofContouringMethods}
Here further details about the creation of the motion adapted sums of contours $S^N$ are given. The motion adapted sum $S_i$ for phase $i$ is calculated with the contour $I_i$ and the deformations of $I_i$ with $T^N$. The following pseudo code shows algorithmically how the motion adapted sums were created:
\begin{itemize}
\item[] for all $i \in N$ do:
\begin{enumerate}[label*=\arabic*)]
\item initialise $S_i$ with $I_i$
\item set $j$ to $i + 1$
\item deform $I_i$ with $T_i^j$ and add result to $S_i$
\item if $j < N - 1$ set $j$ to $j + 1$ otherwise set $j$ to 0 
\item if $j$ is equal to $i$ \textit{continue} with next $i$ otherwise go to 3)
\end{enumerate}
\end{itemize}
The add operation in 3) is performed voxel wise as all $I_n$ in $I^N$ have the same dimension, voxel size and origin. Binary image representation (0:background, 1: foreground) of the contours was used for the calculations.

\section{}
\label{app:doseAnal}
A motion corrected 3DDose volume $M_i$ was created by summing up the dose values in the 4DDose $I^N$ along a voxels trajectory given by the DVF $T^N$, starting from $I_i$ . For every $I_i \in I^N$ one motion corrected dose M i was created:

\begin{itemize}
\item[] for all $i \in N$ do:
\begin{enumerate}[label=\arabic*)]
\item initialise $M_i$ with $I_i$
\item set $j$ to $i + 1$
\item for each $x \in M_i$ do $M_i(x) = M_i(x) + I_j(T_i^j(x))$
\item if $j < N - 1$ set $j$ to $j + 1$ otherwise set $j$ to 0 
\item if $j$ is equal to $i$ \textit{continue} with next $i$ otherwise go to 3)
\end{enumerate}
\end{itemize}

\section{}
\label{app:motionAnal}
In the motion analysis we calculated the maximal motion amplitude $amp_{max}$ for the border region voxels in all phase contours $I_i \in I^N$. The maximal motion extent $amp_{max}$ for a voxel $x$ in time phase $I_i$ was estimated by following $x$ on its trajectory in $T^N$ and tracking its position: 

\begin{enumerate}[label=\arabic*)]
\item set $j$ to $i + 1$ and $amp_{max}$ to 0
\item set $amp$ to $|x - T_i^j(x)|$
\item if $amp > amp_{max}$ set $amp_{max}$ to $amp$
\item if $j < N - 1$ set $j$ to $j + 1$ otherwise set $j$ to 0 
\item if $j$ is equal to $i$ terminate otherwise go to 2)
\end{enumerate}

$amp_{max}$ was calculated separately in x-, y- and z-direction.

\end{document}